\documentclass[conference]{IEEEtran}
\IEEEoverridecommandlockouts

\usepackage{graphicx}
\usepackage{amsmath}
\usepackage{url}
\usepackage{booktabs}
\usepackage{multirow}
\usepackage{listings}
\usepackage{xcolor}
\usepackage{soul}
\sethlcolor{yellow}
\usepackage[utf8]{inputenc} 
\usepackage{comment}

\usepackage{booktabs,tabularx,array,makecell}
\newcolumntype{Y}{>{\raggedright\arraybackslash}X}          
\newcolumntype{C}[1]{>{\centering\arraybackslash}p{#1}}
\usepackage[hidelinks]{hyperref} 
\usepackage{float}
\usepackage{dblfloatfix}   
\newcolumntype{Z}{>{\raggedright\arraybackslash}X}

\begin{document}

\title{Evaluating Endpoint Detection Robustness
Against Genetic Algorithm–Driven Code Transformations}

\title{Evaluating Endpoint Detection Robustness
Against Genetic Algorithm Driven Code Transformations\\}

\author{\IEEEauthorblockN{ Alvina Rwaichi Minja}
\IEEEauthorblockA{\textit{College of Engineering} \\
\textit{ Carnegie Mellon University Africa}\\
Kigali, Rwanda \\
aminja@andrew.cmu.edu}
\and
\IEEEauthorblockN{Jema David Ndibwile}
\IEEEauthorblockA{\textit{College of Engineering} \\
\textit{Carnegie Mellon University Africa}\\
Kigali, Rwanda \\
jndibwil@andrew.cmu.edu}
}

\maketitle

\begin{abstract}
Post-compromise test variants are widely used in controlled security evaluation and endpoint robustness benchmarking. However, modern Antivirus (AV) and Endpoint Detection and Response (EDR) systems increasingly combine signature- and behavior-based detection, challenging the reliability of conventional detection pipelines under adaptive variation. This study introduces ShellForge, a Genetic Algorithm (GA)-driven framework that evolves post-compromise variants representative of remote command execution to generate functionally equivalent variants for systematic detection evaluation. ShellForge applies syntactic transformations, encoding schemes, and structural
permutations guided by a multi-objective fitness function informed by AV and EDR detection feedback. We compare ShellForge against representative baseline transformation frameworks under identical sandbox configurations. Our findings highlight measurable robustness gaps in baseline signature- and behavior-oriented detection pipelines
under controlled variant generation. In addition, we propose a reproducible benchmark for endpoint detection robustness evaluation, motivating the need for robustness-aware defensive monitoring and behavioral correlation.
\end{abstract}

\begin{IEEEkeywords}
Endpoint Detection and Response (EDR), Detection Robustness Benchmarking, Genetic Algorithms, Defensive Robustness Testing, Semantic-Preserving transformations, Post-Compromise Telemetry
\end{IEEEkeywords}

\section{Introduction} 
Modern Endpoint protection platforms increasingly rely on a combination of signature-based antivirus scanning and behavioral telemetry \cite{le_faou_antivirus_2024}. While these defenses have significantly improved, their robustness against adaptive and automatically mutated code variants remains insufficiently understood. This work investigates the detection boundaries of common defensive mechanisms by systematically generating functionally equivalent test variants in a controlled laboratory environment. Such benchmarking supports the development of more resilient signature and behavioral correlation mechanisms, particularly for post-compromise monitoring \cite{traore_red_2024}.

Post-compromise variants are widely used in controlled security evaluations to
study endpoint monitoring and response behaviors \cite{johnson_evading_2021}. Reverse-shell mechanisms are commonly used as representative post-compromise behaviors in controlled security evaluations \cite{kaushik_novel_2021}. The effectiveness of shells and backdoors has been demonstrated in high-profile breaches such as the Hafnium campaign, where operators used outbound callback mechanisms to maintain persistent remote access. Despite their routine usage, systematic evaluation of detection robustness against these post-compromise
variants remains limited.

Recent reports indicate that automatically transformed code variants may lead to
inconsistent responses across endpoint detection layers. This motivates the need
for controlled robustness benchmarking frameworks that help defenders assess
coverage gaps under semantic-preserving transformations
 \cite{mandvi_threat_2025}. 

Obfuscation techniques for payload generation have evolved over the years. Machine Learning (ML) and Large Language Model (LLM)-based approaches are increasingly used to automate payload generation and obfuscation. A growing landscape of \textit{techniques} is used in these concepts, such as LLM-based code generation~\cite{cirkovic_utilizing_2025, khan_ll-xss_2024} and dynamic-programming guided mutation of instructions~\cite{kingful_dynamic_2023}, which show significant improvements in payload generation. However, these approaches remain constrained by \textit{template} dependencies, data constraints, reproducibility concerns, and limited behavioral robustness capabilities. Most current systems generate payloads using established tools like \texttt{MsfVenom, TheFatRat, and Veil}~\cite{kv_av_evasion_2025}. Despite their adoption over the years, these tools often trigger baseline alerts in modern AV/EDR systems since their signatures are known to security vendors~\cite{le_faou_antivirus_2024, mandvi_threat_2025}.

In contrast, the growing field of evolutionary computation suggests great potential for advancing detection-robust transformation strategies. Genetic Algorithm (GA) in particular presents a promising principled approach to systematically evolve polymorphic payloads across multiple generations. By encoding transformations as genetic operators, GA can automatically search for payload variants that preserve functionality while enabling systematic detection robustness evaluation, demonstrating the effectiveness of evolutionary optimization in the generation of novel attack vectors, as shown in GAXSS by Liu \textit{et al.} \cite{liu_gaxss_2022}.

This research aims to evaluate the effectiveness of a GA-driven approach for evolving post-compromise variants representing remote command execution behaviors. These variants are evaluated against both static AV and behavioral EDR detection mechanisms. The question that this research aims to answer is “How effective is a GA-driven approach in generating functionally equivalent post-compromise test variants for evaluating AV/EDR detection robustness?” 

Rather than proposing offensive deployment, ShellForge is designed as a
defensive evaluation framework that enables repeatable testing of how
detection pipelines respond to minor semantic-preserving transformations.
By treating reverse-shell variants as representative post-compromise test
cases, the system supports defenders in understanding which forms of code
mutation are most likely to reduce baseline detection signals and therefore
require improved monitoring.

Our contribution is not the development of an offensive tool but a
systematic measurement study of detection robustness under automatically
generated code variants. All experiments were conducted within controlled
laboratory environments using intentionally generated test variants for
defensive benchmarking purposes. The results highlight practical blind
spots in current defensive systems and motivate the need for stronger
behavioral correlation, anomaly detection, and policy-based controls
beyond static signature matching. Unlike prior work, we formalize payload
evolution as a constrained optimization problem under detection feedback,
where candidate variants are iteratively refined based on detection
responses while preserving functional correctness.

The sections that follow present a review of related work and research methodology, and highlight the anticipated contributions of this study.

\section{Novelty and Main Contributions}
To the best of our knowledge, existing work has not yet examined the use of genetic algorithms to evolve reverse-shell post-compromise test variants under both signature-based antivirus scanning and behavioral EDR-style analysis in a unified framework. Prior studies have typically focused on web payload evolution or binary-only transformations, leaving reverse shells and multi-stage AV/EDR evaluation less explored. ShellForge is intended to help fill this gap by providing a fitness-driven approach to adaptive payload mutation and benchmarking.

The main contributions of this paper are as follows:
\begin{enumerate}
    \item Design of a genetic algorithm--driven framework for generating functionally equivalent post-compromise test variants to support AV/EDR robustness benchmarking.

    \item A multistage fitness evaluation pipeline that integrates both static and
behavioral signals for defensive robustness assessment.

    \item Assessment of GA operators (tournament selection, uniform crossover,
    and random mutation) in evolving payloads across generations.
\end{enumerate} 


\section{Literature Review} 
Automated payload generation is not a new concept in cybersecurity operations. Since 2003 tools such as Metasploit have been used as platforms for penetration testing, offering a wide range of exploits and payloads on security vulnerabilities \cite{kennedy_brief_2011}. Over time, the concept has expanded beyond simply generating executable test samples to producing lower-alerting, encoded variants for defensive testing. Several tools integrated within Kali Linux and Metasploit, including MsfVenom, TheFatRat, Unicorn, Veil show this progression by applying various layers of obfuscation \cite{rapid7_msfvenom_docs, thefatrat_github, trustedsec_unicorn,kali_veil_tool}. A study by Kiran \textit{et al}. evaluating payload obfuscation tools highlights this evolution, showing the effectiveness of frameworks such as Veil \cite{kv_av_evasion_2025}. However, the same study emphasizes the need for improvement in achieving both behavioral and static detection robustness, as well as in developing automation to meet modern red-team requirements.

Recent work has explored machine-learning and generative
approaches for automated payload generation and obfuscation.
Examples include LLM-based code generation and
dynamic-programming–guided instruction mutation. Cirkovic \textit{et al.} used LLMs to generate synthetic polymorphic payloads for Cross-site Scripting (XSS), SQL Injections, and Command Injections vulnerabilities \cite{cirkovic_utilizing_2025}. Their work demonstrated the growing role of generative methods in automated security testing and defensive robustness evaluation. However, the approach was limited to web application payloads and heavily dependent on the training dataset, and did not explicitly optimize detection robustness against AV or EDR tools.  Khan \textit{et al.} \cite{khan_ll-xss_2024} similarly used LLMs for XSS payloads but did not extend the evaluation to behavioral defenses, leaving their work constrained to static robustness.

In contrast, Kingful \textit{et al.} used Dynamic Programming (DP) as a reinforcement-learning–inspired optimization approach to generate adversarial Windows payloads. Their approach utilized three transformations; NOP (No Operation) Insertions, insertion of jump instructions and replacement of instructions to achieve obfuscation of payloads \cite{kingful_dynamic_2023}. Although their approach guaranteed functional payloads with minimal overhead, they had limited evaluations and overlooked behavioral defenses. Furthermore, DP’s deterministic nature narrowed the search space, producing similar payloads that defenders could patch against. As a result, while their study validated the feasibility of adversarial optimization, it also shows the need for stochastic, population-based approaches capable of richer transformations and cross-platform applicability, balancing both static and behavioral robustness.
Liu \textit{et al}. have explored evolutionary algorithms, specifically GA in the domain of XSS vulnerabilities \cite{liu_gaxss_2022}. Their study effectively demonstrated the capacity of GA to explore large, complex search spaces for adversarial generation. They achieved high detection rates, free dictionary operation, and adaptability across multiple applications. However, their focus was primarily on XSS vulnerability detection rather than stealth or evasion. Other recent studies across diverse domains highlight the importance of robust evaluation methodologies under varying conditions \cite{demirci2015knowledge} . This demonstrates a promising opportunity to expand the application of GAs beyond vulnerability discovery to stealth optimization in payload generation, particularly for post-compromise variants.

With a focus on detection robustness, it is important to distinguish between the wide range of detection evasion techniques described in practice. Le Faou \cite{le_faou_antivirus_2024} provides a summary of common techniques employed against AV/EDR solutions, from string obfuscation and packing to in-memory execution and API unhooking. However, as the reviewed literature above shows, most research contributions have concentrated on defeating static detection mechanisms, with limited evaluation against behavioral or dynamic defenses. 

To situate this project within that context, \autoref{tab:evasion-comparison} presents a structured comparison of evasion tactics across existing studies, pointing to the unexplored potential of Genetic Algorithms for post-compromise variant obfuscation within the proposed ShellForge framework.

\begin{table}[H]
\centering
\caption{Transformation Technique Comparison}
\label{tab:evasion-comparison}
\begin{tabularx}{\columnwidth}{@{} l c X X @{}}
\toprule
\textbf{Paper} & \textbf{Static} & \textbf{Behavioral} & \textbf{Testing Tool} \\
\midrule
Cirkovic \textit{et al}. (2025)      & Yes & No      & Damn Vulnerable Web Application (DVWA) \\
Kingful \textit{et al}. (2023)       & Yes & Partial (checked functionality of payload in sandbox but not evasion) & EMBER, ClamAV, VirusTotal, Cuckoo sandbox \\
Liu \textit{et al}. (2022) GAXSS     & Yes & No      & 6 open-source web applications \\
Khan (2024) LL-XSS          & Yes & No      & OWASP Juice Shop \\
Li \textit{et al}. (2024)            & Yes & No      & Metasploit / MsfVenom (UI) \\
Song \textit{et al}. (2022)          & Yes & Partial & MalConv, Ember-style models, commercial AV engines \\
Demetrio \textit{et al}. (2021)      & Yes & Partial (functionality test only) & EMBER, MalConv, DNNs, GBT, VirusTotal \\
Johnson \& Haddad (2021)    & Yes & No      & VirusTotal, commercial AVs \\
\textit{ShellForge}         & Yes & Yes     & VirusTotal, commercial AVs, Cuckoo sandbox, EDR telemetry \\
\bottomrule
\end{tabularx}
\end{table}

 A key research gap lies in applying genetic algorithms to the evolution of post-compromise execution variants. From the reviewed literature, no study simultaneously evaluates both static and behavioral detection mechanisms while employing genetic algorithms specifically for executable samples executed by remote commands. \autoref{tab:payload-methods-wide} shows a comparison of the existing payload generation techniques and payload types used in prior work, highlighting the absence of GA-driven optimization for reverse shell.

\begin{table*}[!ht]
\centering
\caption{Comparison of Payload Generation Methods and Payload Types}
\label{tab:payload-methods-wide}
\begin{tabularx}{\textwidth}{@{} l l Y Y l @{}}
\toprule
\textbf{Paper} & \textbf{Payload Type} & \textbf{Generation} & \textbf{Core Algorithm/Technique} & \textbf{Algorithm} \\
\midrule
Cirkovic \textit{et al}. (2025) & Web (XSS, SQLi) & Automated payload text generation & LLM fine-tuning & Neural networks \\
Kingful \textit{et al}. (2023) & Windows PE & Functionality-preserving binary transforms & Code transformations & Dynamic programming \\
Liu \textit{et al}. (2022) GAXSS & Web (XSS) & Evolved attack vectors (strings) & Vector evolution & Genetic algorithm \\
Khan (2024) LL-XSS & Web (XSS) & Context-aware XSS generation & Generative models & Neural networks \\
Li \textit{et al}. (2024) & Multiple (Metasploit payloads) & GUI wrapper for \texttt{MsfVenom} payloads & GUI wrapper & None (tool interface) \\
Song \textit{et al}. (2022) & Windows malware & Reward system & Binary modification & Reinforcement learning \\
Demetrio \textit{et al}. (2021) & Windows PE & Functionality-preserving & Header manipulation & RAMEN: white-box; GA black-box \\
Johnson \& Haddad (2021) & Reverse shell (custom C exe) & Manual staging of payloads & HTTP staging & Manual modification \\
\textit{ShellForge} & Reverse shell & Evolving genomes & Evolutionary generation & Genetic algorithm \\
\bottomrule
\end{tabularx}
\end{table*}

\newcolumntype{L}[1]{>{\raggedright\arraybackslash}p{#1}}
\newcolumntype{C}[1]{>{\centering\arraybackslash}p{#1}}

It is also important to highlight the gap in cross-platform applicability. As shown in \autoref{tab:platform-language}, prior works have primarily focused either on Windows PE or web-based implementations, with little attention to other payload execution formats. ShellForge has strong potential to address this limitation, as GA can be structured to generate and evolve payloads across multiple formats, thereby extending applicability beyond the narrow scope of existing studies.

\newcolumntype{L}[1]{>{\raggedright\arraybackslash}p{#1}}
\newcolumntype{C}[1]{>{\centering\arraybackslash}p{#1}}

\begin{table}[!ht]
\centering
\setlength{\tabcolsep}{3pt}           
\renewcommand{\arraystretch}{1.5}    
\footnotesize
\caption{Platform and Language Coverage}
\label{tab:platform-language}
\begin{tabular}{@{} L{0.35\columnwidth} C{0.18\columnwidth} C{0.18\columnwidth} L{0.18\columnwidth} @{}}
\toprule
\textbf{Paper} & \textbf{Windows PE shell code} & \textbf{\makecell{(ELF /\\ Scripts)}} & \textbf{Language} \\
\midrule
Cirkovic \textit{et al}. (2025)        & No  & No  & Web languages \\
Kingful \textit{et al}. (2023)         & Yes & No  & Windows-specific \\
Liu \textit{et al}. (2022) GAXSS       & No  & No  & JavaScript, HTML \\
Khan (2024) LL-XSS            & No  & No  & JavaScript, HTML \\
Song \textit{et al}. (2022)            & Yes & No  & Windows binary \\
Demetrio \textit{et al}. (2021)        & Yes & No  & Windows PE headers \\
Johnson \& Haddad (2021)      & Yes & No  & C/C++ \\
\textit{ShellForge} & Yes & Yes & Multiple \\
\bottomrule
\end{tabular}
\end{table}

\section{Methodology}

This section describes the methodology used to design and evaluate
\textit{ShellForge}, a Genetic Algorithm (GA)-driven framework for generating security test variants, including reverse-shell execution samples, and evaluating detection robustness under automatically mutated variants in a secure and controlled environment.
The methodology covers the experimental setup, toolchain, and payload
transformation design, GA configuration, and evaluation protocol.
The objective is to generate reverse-shell variants that preserve
operational behavior while producing different detection responses
under controlled evaluation by modern Antivirus (AV) and
Endpoint Detection and Response (EDR) systems.

\subsection{Lab Setup and Environment Configuration}
All experiments were conducted in an isolated virtual testbed to ensure
safety, reproducibility, and controlled execution conditions.

\subsubsection{Host System Specifications}
The host machine used for orchestration and payload generation was
configured as follows:

\begin{itemize}
    \item Operating system: Windows 11 Pro
    \item Processor: 13\textsuperscript{th} Gen Intel Core i7-1355U
    \item Virtualization platform: VirtualBox 7.1.4
\end{itemize}

\subsubsection{Virtual Machine Configuration}
Three virtual machines were deployed to separate defensive robustness testing,
target execution, and behavioral analysis.

\textbf{VM 1: Generation and Orchestration Machine (Kali Linux).}
This machine was used for payload generation and listener deployment.

\begin{itemize}
    \item OS: Kali Linux 2025.2
    \item Resources: 4\,GB RAM, 2 CPU cores
    \item Network mode: Host-only adapter
\end{itemize}

\textbf{VM 2: Target Machine (Windows 11).}
This environment executed candidate payloads and enabled local AV testing.

\begin{itemize}
    \item OS: Windows 11 Professional
    \item Resources: 4\,GB RAM, 2 CPU cores
    \item Security engines: Windows Defender, ClamAV
    \item Network mode: Host-only adapter
\end{itemize}

\textbf{VM 3: Analysis Machine (CAPE Sandbox).}
This machine supported behavioral inspection and sandbox telemetry
collection.

\begin{itemize}
    \item OS: Ubuntu 22.04 LTS Server
    \item Resources: 16\,GB RAM, 50\,GB disk
    \item Sandbox platform: CAPE/Cuckoo Sandbox
    \item Network mode: Bridged and Host-only adapters
\end{itemize}

\subsection{Tools and Frameworks}

\subsubsection{Test Artifact Generation Tools}
Payload variants were generated and transformed using the following toolchains:

\begin{itemize}
    \item Python 3.x: primary implementation language for payload encoding
    \item \texttt{MsfVenom}: baseline reverse-shell payload generator
    \item Veil Framework: reference baseline transformation framework for comparison

\end{itemize}

\subsubsection{Genetic Algorithm Implementation}
ShellForge employs a GA to evolve payload transformation strategies.
The implementation is based on the \texttt{DEAP} library (Distributed
Evolutionary Algorithms in Python).

\paragraph{Chromosome Representation}
Each chromosome encodes a complete payload-generation strategy,
including the selected obfuscation operators, their ordering, and
associated parameter values.

\paragraph{Fitness Function}
Candidate payloads are optimized using a multi-objective fitness score function as shown in \eqref{eq:fitness}
balancing static robustness, behavioral robustness, functionality, and efficiency:

\begin{equation}
\mathcal{F} = w_sE_s + w_bE_b + w_fC + w_\eta\eta,
\quad
w_s + w_b + w_f + w_\eta = 1,
\label{eq:fitness}
\end{equation}

where $E_s$ denotes resistance to signature-based AV detection,
$E_b$ captures behavioral response under sandbox-based analysis and runtime analysis,
$C$ represents functional correctness (successful reverse-shell execution),
and $\eta$ measures payload efficiency in terms of size and execution
overhead.

\paragraph{Evolutionary Operators}
ShellForge applies standard GA operators:

\begin{itemize}
    \item Selection: tournament selection with $k=3$
    \item Crossover: uniform crossover with probability $p_c=0.7$
    \item Mutation: random mutation with probability $p_m=0.2$
\end{itemize}

\paragraph{GA Parameters}
The evolutionary process was configured as follows:

\begin{itemize}
    \item Population size: $N=25$
    \item Number of generations: 20--30 (or until convergence)
    \item Early stopping: termination after $g_{\text{stall}}=5$ stagnant generations
\end{itemize}

\subsubsection{Detection and Analysis Tools}
Static and behavioral evaluations were performed using:

\begin{itemize}
    \item VirusTotal: multi-engine AV scanning (static analysis)
    \item ClamAV: open-source signature-based detection
    \item Windows Defender: native endpoint protection testing
    \item Wireshark: network traffic monitoring and validation
\end{itemize}

\subsection{Payload Transformation Design}
Each GA chromosome specifies a transformation pipeline applied to a base
reverse shell generated via \texttt{MsfVenom}
(\texttt{python/shell\_reverse\_tcp}).
The GA explores combinations of obfuscation operators, including:

\begin{itemize}
    \item XOR encoding
    \item Base64 wrapping
    \item ROT-based encoding
    \item String splitting and reconstruction
    \item Variable and function renaming
    \item Dynamic imports and symbol resolution
    \item Dead-code insertion
    \item Analysis-aware runtime checks
      \item Polymorphic encoding strategies
\end{itemize}

These transformations produce polymorphic payload variants while ensuring
syntactic correctness and operational validity.

\subsection{Testing and Evaluation Protocol}
Each generated payload was evaluated according to four weighted criteria:

\begin{itemize}
    \item \textbf{Static robustness score (40\%):} detection resistance against
    VirusTotal, ClamAV, and Windows Defender.
    
    \item \textbf{Behavioral robustness score (35\%):} runtime stealth assessed
    through CAPE/Cuckoo sandbox telemetry, Sysmon monitoring, and network
    traces in Wireshark.
    
    \item \textbf{Functionality (20\%):} successful reverse-shell callback
    establishment and session stability.
    
    \item \textbf{Efficiency (5\%):} payload compactness and execution overhead.
\end{itemize}

GA-evolved variants were benchmarked against baseline frameworks under
identical evaluation conditions, including unencoded payloads,
\texttt{MsfVenom} with the \texttt{shikata\_ga\_nai} encoder, and Veil
Framework Python-based evasion techniques.


\section{System Architecture}

\begin{figure}[!t]
  \centering
  \includegraphics[width=\columnwidth]{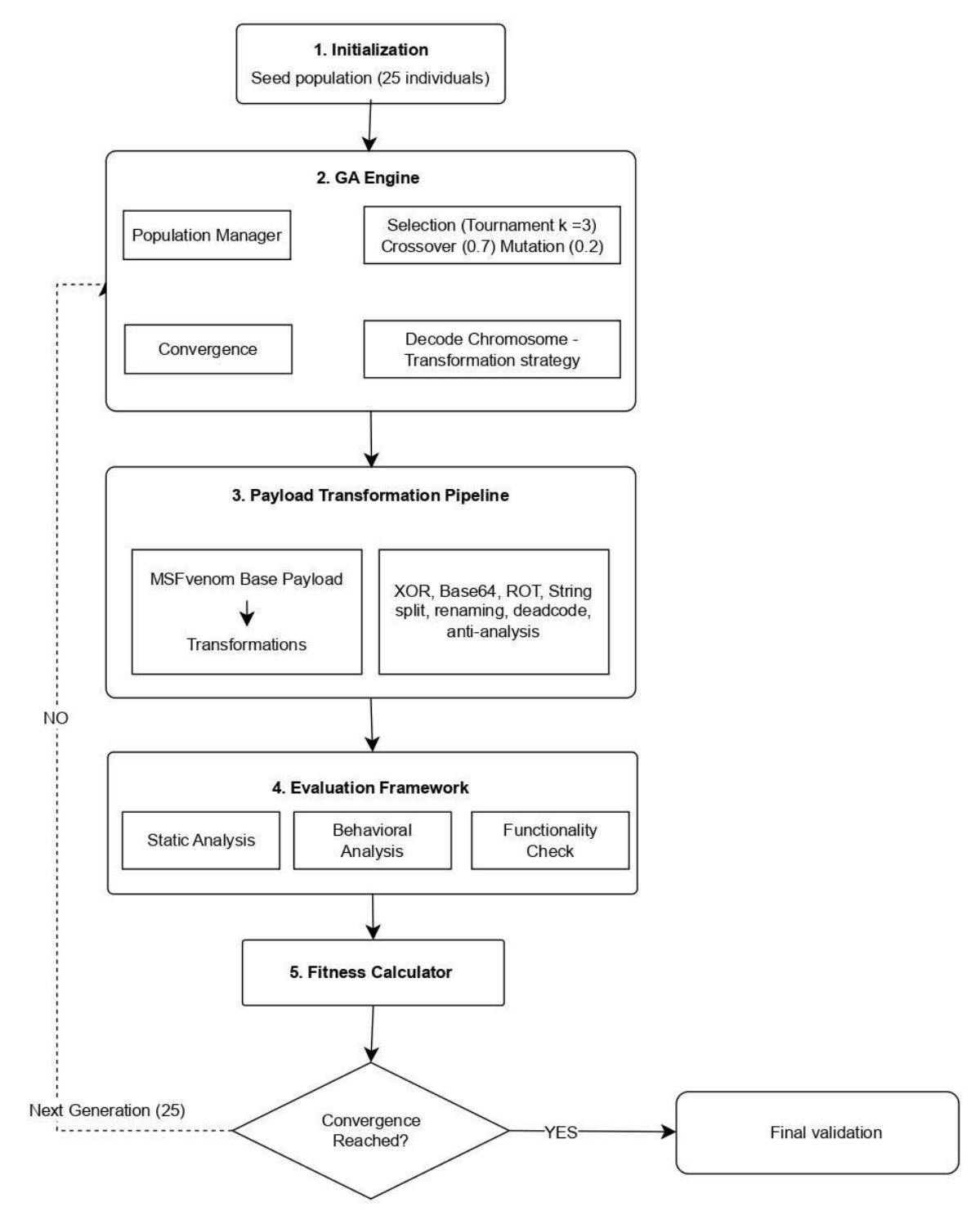}
  \caption{ShellForge System Architecture}
  \label{fig:shellforge_architecture}
\end{figure}

This section presents a clear description of the \textit{ShellForge} system architecture. The system comprises three integrated subsystems: the Genetic Algorithm (GA) Engine, a Payload Transformation Pipeline, and an Evaluation Framework. Together, they operate as a closed feedback loop to iteratively evolve, test, and benchmark payload variants derived from canonical templates. The interaction among these components is illustrated in  Figure~\ref{fig:shellforge_architecture}.


\subsection{Genetic Algorithm Engine}
The GA engine manages and guides the evolutionary search process. Its key components include: the \textit{Population Manager}, which maintains the current population and provisions initial populations drawn from canonical templates such as MsfVenom payloads; The \textit{Evolution Controller}, which implements selection, recombination and mutation operators; and the \textit{Convergence Monitor}, which observes generational fitness trajectories and triggers termination when improvement falls below a minimal threshold or when maximum generation count is reached.

The selection policy uses tournament selection ($k = 3$), recombination uses uniform crossover (rate = 0.7), and mutation is applied stochastically (rate = 0.2). The engine implements elitism by preserving the top 10\% of high-fitness individuals between generations, stabilizing performance, and accelerating convergence.

\subsection{Payload Transformation Pipeline}
The transformation pipeline maps GA chromosomes to transformation-based payload variants while ensuring functional integrity. 
The \textit{Chromosome Decoder} interprets each chromosome as an ordered sequence of transformation operators with associated parameters. 
The \textit{Transformation Engine} applies the decoded operators to payload templates to produce obfuscated variants. 
Finally, the \textit{Payload Assembler} merges transformed fragments into a valid executable payload, maintaining syntactic correctness to ensure operational viability.

\subsection{Evaluation Framework}
This framework provides measured fitness values by exercising candidate payloads through automated static and behavioral tests. The \textit{Static Analysis Module} aggregates detection metrics from multiple static sources like VirusTotal API and local signature based engines like ClamAV. The \textit{Behavioral Analysis Module} executes payloads in an isolated environment and collects runtime process activity, file/system interactions and network traces to identify dynamic detections. The \textit{Functionality Check} confirms that each of the transformed payloads achieves the intended functionality. Finally,  the \textit{Fitness Calculator} aggregates component scores into a composite fitness value according to the defined criteria.

These subsystems operate in a continuous optimization cycle: the GA Engine evolves payload strategies, the Transformation Pipeline applies these strategies to generate variants, and the Evaluation Framework scores each variant's performance. Fitness scores feed back into the GA Engine to guide the next generation, creating an iterative improvement process.
This architecture provides a controlled and reproducible environment for investigating how evolutionary code transformations affect both detection response and payload functionality. It serves as the experimental foundation for the implementation and analysis phases that follow.

\section{Implementation}

ShellForge is implemented as an integrated evolutionary framework that
jointly optimizes static obfuscation and behavioral robustness for reverse-shell automated variants. The system encapsulates the Genetic Algorithm (GA), payload transformation modules, and a multi-stage evaluation pipeline into a coordinated architecture that supports reproducible and iterative payload evolution.

\subsection{System Components}
ShellForge consists of seven tightly coupled modules, each responsible for a specific stage in the evolutionary workflow:

\begin{itemize}
    \item \textbf{Configuration Manager:} defines global experiment parameters, GA hyperparameters and transformation constraints.

    \item \textbf{Chromosome Encoder:} provides a unified genome representation
    encoding both static transformations and behavioral robustness parameters.

    \item \textbf{Transformation Engine:} implements the static obfuscation library, including XOR, Base64, and ROT encodings, string splitting,
    renaming, and structural mutations.

    \item \textbf{Behavioral Tactics Module:} implements runtime transformation-aware techniques such as VM probing, resource checks, delays, and
    analysis-aware runtime checks.

    \item \textbf{Fitness Evaluator:} executes the multi-stage testing pipeline, combining static AV scanning, sandbox telemetry, network traces, and
    functionality validation.

    \item \textbf{GA Engine:} performs evolutionary optimization using tournament selection, uniform crossover, mutation, and elitism.

    \item \textbf{System Orchestrator:} coordinates data flow across modules, schedules evaluation cycles, maintains generational state, and manages output variants.
\end{itemize}

Together, these components enable ShellForge to explore a large transformation space, generate syntactically valid payload variants, and converge toward high-robustness solutions across successive generations.

\subsection{Chromosome Representation}
\begin{figure}[!t]
\centering
\includegraphics[width=\columnwidth]{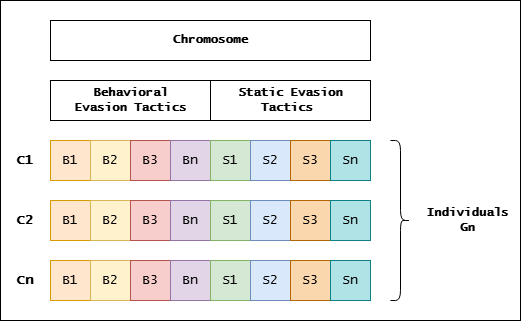}
\caption{Unified chromosome encoding behavioral parameters $(B_1,\dots,B_n)$
and static transformation operators $(S_1,\dots,S_n)$.}
\label{fig:chromosome-structure}
\end{figure}

\autoref{fig:chromosome-structure} illustrates the unified chromosome design. Each chromosome encodes a complete transformation strategy through two gene segments:
behavioral parameters $(B_1,\dots,B_n)$ and static transformations
$(S_1,\dots,S_n)$.

The behavioral segment contains evolvable runtime parameters, including
execution delays (5--90~s), resource thresholds (RAM: 512--8192~MB,
CPU: 1--8 cores, uptime: 1--30~min), and boolean flags controlling environment
checks (VM detection, debugger probing, DNS validation, and artifact scanning).
The static segment encodes an ordered transformation pipeline (e.g.,
XOR encoding $\rightarrow$ Base64 wrapping), where both operator selection and ordering are subject to evolution.

This dual representation enables simultaneous optimization across static and behavioral robustness spaces, with each individual representing a distinct payload variant refined through iterative selection.

\subsection{Multi-Stage Fitness Evaluation}
Each candidate payload undergoes a structured evaluation pipeline designed to measure both detection robustness and operational correctness.

\subsubsection{Stage 1: Functionality Validation}
In our experiments, we instantiate reverse-shell behaviors as a controlled test case for post-compromise activity. A reverse-shell tester spawns a listener on port~4444, executes the candidate payload, confirms TCP callback establishment, verifies command execution (e.g., \texttt{whoami}, \texttt{pwd}), and assesses session
stability. Payloads that fail the functionality check receive a score of zero and are discarded from further evaluation.

\subsubsection{Stage 2: Local Static Detection Testing}
Static signature-based detection is performed using ClamAV and Windows Defender. ClamAV executes lightweight scans in the Linux environment, producing a binary outcome (detected or clean). In parallel, Windows Defender scans the payload within the target Windows virtual machine using \texttt{Start-MpScan}.

The combined local static score is computed as depicted in \eqref{eq:Slocal}:

\begin{equation}
S_{\text{local}} = \frac{2-\text{detections}}{2},
\label{eq:Slocal}
\end{equation}

where \textit{detections} denotes the number of engines (0--2) flagging the
payload.

\subsubsection{Stage 3: VirusTotal Multi-Engine Analysis}
To incorporate broader commercial coverage, only the top 20\% of candidates proceed to VirusTotal scanning (respecting API rate limits), where we computed the Virustotal detection scores as shown in \eqref{eq:Scorevt}:

\begin{equation}
S_{\text{vt}} = \frac{62-\text{detections}}{62}.
\label{eq:Scorevt}
\end{equation}

The static robustness score is defined as shown in \eqref{eq:staticscore}:

\begin{equation}
S_{\text{static}} = 0.3\,S_{\text{local}} + 0.7\,S_{\text{vt}}.
\label{eq:staticscore}
\end{equation}

\subsubsection{Stage 4: Behavioral Sandbox Evaluation}
Behavioral characteristics are assessed through CAPE sandbox execution with a maximum runtime of 120~s. The behavioral score incorporates weighted penalties and rewards across multiple runtime indicators:

\begin{itemize}
    \item Signature triggers (40\%)
    \item Execution duration (25\%)
    \item Process tree complexity (15\%)
    \item Network activity profile (10\%)
    \item File-system operations (5\%)
    \item Command execution artifacts (5\%)
\end{itemize}

If CAPE is unavailable, a fallback behavioral score of 0.85 is assigned.

\subsubsection{Stage 5: Efficiency Metric}
Payload efficiency is incorporated as a lightweight penalty on payload size as seen in \eqref{eq:efficiency}:

\begin{equation}
S_{\text{efficiency}} =
\max\left(0.1,\frac{500}{\text{payload\_size\_bytes}}\right).
\label{eq:efficiency}
\end{equation}

This rewards compact payloads while penalizing excessive wrapper overhead.

\subsection{Total Fitness Score}
The overall fitness is computed as a weighted aggregation as depicted in \eqref{eq:totalfitness}:

\begin{equation}
\begin{aligned}
F_{total} = & \; 0.40\,S_{static} + 0.35\,S_{behavioral} \\
            & + 0.20\,S_{functionality} + 0.05\,S_{efficiency}
\end{aligned}
\label{eq:totalfitness}
\end{equation}

Here, $S_{\text{functionality}}$ is binary, while static and behavioral scores
are continuous measures derived from AVs and sandbox.

\subsection{Evolution Workflow}
At each generation, ShellForge applies tournament selection, uniform crossover,
and stochastic mutation to produce a new population, while preserving top
individuals through elitism. Validity checks ensure that all evolved payloads
remain executable.

To reduce computational overhead, SHA256-based caching avoids re-evaluating
identical payloads, and progressive filtering ensures that only high-performing
candidates proceed to expensive evaluation stages such as VirusTotal and CAPE.
Across typical experiments, ShellForge operates with populations of 20
individuals for approximately 15 generations, terminating early when fitness
convergence is detected.

\section{Experimental Results}

This section presents the experimental performance of ShellForge, focusing on
the evolutionary convergence behavior and the characteristics of the best
evolved payload. Results are reported in terms of fitness progression,
robustness outcome, and operational correctness.

\subsection{Evolution Performance}

The evolutionary process converged rapidly, and the best individual reached a fitness score of 0.906 early in the optimization process and remained stable throughout subsequent generations. Convergence was detected by generation~3, where fitness improvement fell below 0.1\% for two consecutive generations. Consequently, the evolutionary process terminated early at generation~5. Table IV summarizes the comparative AV detection outcomes across representative payload-generation frameworks.

\begin{table}[H]
\caption{Comparative AV Detection Results across Frameworks}
\label{tab:comparison}
\centering
\small
\begin{tabular}{lcc}
\hline
\textbf{Framework} & \textbf{VT Detection} & \textbf{Sandbox Verdict} \\
\hline
MsfVenom reverse & 19/62 & Malicious \\
MsfVenom (shikata\_ga\_nai) & 44/62 & Malicious \\
Veil-Evasion & 2/62 & Malicious \\
TheFatRat & 27/62 & Malicious \\
ShellForge (ours) & 0/62 & Inconclusive \\
\hline
\end{tabular}
\label{tab:vt_comparison}
{\footnotesize Detections reported over 62 VirusTotal engines.} 
\end{table}

To further interpret the optimal solution, the fitness components of the best
individual at generation~5 are summarized as follows:

\begin{itemize}
    \item Static robustness score: 1.00 (no alerts were observed under the local signature-scanning configuration used in the experiment)
    \item Behavioral robustness score: 0.85
    \item Functionality score: 1.00 (fully operational reverse shell)
    \item Efficiency score: 0.166 (compact encoding overhead)
    \item Total weighted fitness: 0.906
\end{itemize}

Overall, the results indicate stable evolutionary convergence and rapid discovery of a robust local optimum, demonstrating the effectiveness of GA-driven payload evolution.

\subsection{Best Evolved Payload Performance}

The GA converged to an integrated transformation strategy combining both behavioral
simplicity and strong static obfuscation. The best-performing configuration
was obtained at Generation~5.

\subsubsection{Evolved Behavioral Parameters}

The behavioral gene segment converged toward minimal runtime complexity,
with the following parameterization:

\begin{itemize}
    \item Execution delay: 30--32~s
    \item RAM threshold: 961~MB
    \item CPU requirement: 2 cores
    \item Minimum uptime: 6~min
    \item Environmental checks: disabled (VM, debugger, DNS, artifact probing)
\end{itemize}

\subsubsection{Evolved Static Transformation Strategy}

The static transformation pipeline selected by evolution consisted of:

\begin{itemize}
    \item XOR encoding (key = 42)
    \item Base64 wrapping (single iteration)
\end{itemize}

This combination emerged as the most effective encoding sequence for reducing
signature-based detection while preserving payload validity.
\subsubsection{Functionality Validation}


To confirm that detection robustness was achieved without sacrificing operational correctness, each payload variant was subjected to a functionality validation workflow while preserving the same reverse-shell semantics. Table~\ref{tab:functionality} reports successful connection establishment, command execution, and session stability.

\begin{table}[!htbp]
\centering
\caption{Functionality Validation of the Best Evolved Payload}
\label{tab:functionality}
\renewcommand{\arraystretch}{1.15}
\setlength{\tabcolsep}{4pt}
\begin{tabular}{l c}
\toprule
\textbf{Test} & \textbf{Result} \\
\midrule
Connection establishment & Successful \\
Command execution (\texttt{whoami}, \texttt{pwd}, \texttt{ls}) & Successful \\
Interactive shell stability & Stable \\
Overall functionality & Pass \\
\bottomrule
\end{tabular}
\end{table}

\subsubsection{Payload Characteristics}

The best evolved payload exhibited the following properties:

\begin{itemize}
    \item File size: 3,016 bytes (baseline: $\sim$450 bytes)
    \item Total fitness: 0.906
    \item Static score: 1.00
    \item Behavioral score: 0.85
    \item Functionality score: 1.00
    \item Efficiency score: 0.166
\end{itemize}

\subsubsection{Key Observation}

A key finding is that evolution discovered an effective balance between
behavioral simplicity and static transformation strength. Despite disabling all environmental checks, the evolved configuration reduced baseline alerts under the tested configuration through robust static obfuscation alone. These results suggest that, under signature-based detection, transformation quality exerts a stronger influence on baseline alerting outcome than behavioral complexity.

\section{DISCUSSION}
This section interprets the experimental results and what they imply for robustness-oriented variant generation and defensive evaluation. We discuss which fitness components drove performance and how ShellForge compares with template-based frameworks such as \texttt{MsfVenom}, Veil, and TheFatRat. We also outline key limitations of the current evaluation.

\subsection{Key Findings}

ShellForge produced Python reverse-shell variants that in several configurations did not generate baseline alerts under the tested configuration while preserving functionality. The Genetic Algorithm (GA) converged rapidly: the high-performing configuration appeared in the
first generation and remained stable throughout subsequent iterations,
indicating a well-shaped search space and an effective fitness function.

Relatively small evolutionary populations were used to reflect practical
resource-constrained evaluation scenarios typical in defensive security
testing environments.

A central finding is that minimal behavioral complexity combined with
strong static transformations yields the strongest robustness results.

\subsection{Comparative Evaluation}
We used VirusTotal and behavioral sandbox analysis to compare ShellForge's performance to representative template-based frameworks such as MsfVenom, Veil-Evasion, and TheFatRat under identical assessment conditions. Table IV shows that the ShellForge-generated variant had the lowest VirusTotal detection rate of all the evaluated frameworks. Baseline frameworks produced working payloads but triggered multiple antivirus detections. These findings imply that, in contrast to fixed-template payload generation techniques, automated evolutionary search can generate transformation strategies that result in distinct detection responses.

\subsection{Novelty and Contributions}

ShellForge introduces several contributions not present in existing payload-generation tools:

\begin{itemize}
    \item \textbf{Unified static + behavioral evolution:} Unlike traditional frameworks that apply fixed transformation templates, ShellForge evolves both static encodings and behavioral parameters jointly, enabling a more comprehensive robustness benchmarking strategy.

    \item \textbf{Automated discovery of novel transformations:} The evolved XOR + Base64 combination was not part of any preset obfuscation library. Its emergence through GA search demonstrates that automated optimization can produce functionally equivalent test variants that may trigger lower baseline alerting, supporting defensive evaluation.

    \item \textbf{Lightweight yet effective payloads:} ShellForge produces Python payloads that remain compact while exhibiting different detection outcomes compared to representative baseline frameworks under identical experimental conditions in both static and behavioral detection.

    \item \textbf{Rapid convergence and operational efficiency:} Achieving optimal detection robustness in the first generation shows that the fitness function and operator design efficiently guide the search, enabling fast adaptation.
\end{itemize}

Together, these contributions demonstrate that evolutionary computation can complement template-based approaches by enabling robustness-aware defensive evaluation under semantic-preserving transformations.

\subsection{Implications for Detection}

The comparative study highlights limitations in traditional signature-based detection. All template-driven frameworks (MsfVenom, Veil-Evasion, and TheFatRat) were detected by a majority of VirusTotal engines and consistently flagged as malicious by behavioral sandboxes. In contrast, ShellForge produced a lightweight Python payload with a transformation profile not represented in current signature databases, reducing baseline static alerts in this controlled setting and yielding lower-confidence or inconclusive sandbox verdicts.

These findings suggest a pressing need for detection systems to incorporate semantic code analysis, behavioral modeling, and ML-based classifiers. Reliance on static signatures alone is insufficient when automated variant generation is considered in robustness-oriented defensive testing. In this context, extending the evaluation to heterogeneous execution categories and
cross-language transformation spaces represent a key next step toward
understanding the generalizability of evolutionary transformation-based
testing frameworks.

\subsection{Ethical Considerations}
All experiments in this study were conducted strictly in isolated virtual environments and are intended solely for defensive research and endpoint robustness benchmarking purposes. No real-world infrastructure, unauthorized access, or public deployment was involved. The goal is to inform the design of more robust AV/EDR detection strategies and contribute to safer endpoint security practices. 

\subsection{Limitations}

The experiments in this study were conducted in a controlled laboratory
environment using a limited set of antivirus engines and sandbox
configurations. Detection outcomes may vary across enterprise deployments,
different AV vendors, and evolving signature databases. In addition,
the evaluation focuses on Python-based reverse-shell artifacts used as
representative post-compromise behaviors. While this enables controlled and reproducible
benchmarking of semantic-preserving transformation strategies, it constitutes an initial phase of a broader, extensible evaluation framework. Ongoing work extends this framework toward a multi-category evaluation
setting, incorporating diverse execution types, programming languages,
and behavioral profiles, enabling assessment of  generalization beyond the current scope.

\section{CONCLUSION AND FUTURE WORK}
This section provides an overview of ShellForge's main contributions and findings and highlights the takeaways from the assessment. We then describe the important next steps required to further the framework beyond Python reverse shells, strengthen behavioral evaluation, and improve reproducibility.
\subsection{Conclusion}

This work presented ShellForge, a genetic algorithm framework for evolving functionally equivalent post-compromise test artifacts for defensive robustness benchmarking through unified optimization of static and behavioral features. A key insight from this study is that the robustness outcomes are driven primarily by the strength of static transformation strategies rather than extensive behavioral complexity. Across iterations, the GA converged toward compact and effective obfuscation sequences while minimizing reliance on environmental checks, and the evolved payloads consistently preserved functionality across generations. In this controlled setting, several generated variants triggered fewer baseline
alerts under static scanning. In some configurations, no alerts were raised,
highlighting practical limitations of purely signature-oriented detection when
faced with semantic-preserving transformations. These findings motivate the
need for robustness-aware evaluation methodologies that complement static
matching with behavioral correlation. They further demonstrate the value of automated evolutionary search in discovering transformation patterns that extend beyond the capabilities of fixed-template transformation tools. ShellForge therefore provides a practical demonstration of how adaptive, search-based approaches can meaningfully advance the study of detection robustness under adaptive, semantic-preserving transformations.

\subsection{Future Work}

Future work will extend ShellForge as a defensive benchmarking framework across additional post-compromise behaviors and execution formats,
strengthening behavioral telemetry, and improving reproducibility across endpoint environments. We also plan to support curated evaluation datasets and
robustness-aware testing methodologies that can help defenders assess and
improve detection performance.

\bibliographystyle{IEEEtran}
\bibliography{refs_new}
\end{document}